# Ultra-broadband dual-comb spectroscopy across 1.0–1.9 µm

Sho Okubo,[1,3†*] Kana Iwakuni,[1,2,3†] Hajime Inaba,[1,3] Kazumoto Hosaka,[1,3] Atsushi Onae,[1,3] Hiroyuki Sasada,[2,3] and Feng-Lei Hong[1,3]

[1]*National Metrology Institute of Japan (NMIJ), National Institute of Advanced Industrial Science and Technology (AIST), Tsukuba, Ibaraki 305-8563, Japan*

[2]*Department of Physics, Faculty of Science and Technology, Keio University, Kohoku, Yokohama 223-8522, Japan*

[3]*JST, ERATO, MINOSHIMA Intelligent Optical Synthesizer Project, Tsukuba, Ibaraki 305-8563, Japan*

[†]These authors contributed equally to this work.

[*]E-mail: sho-ookubo@aist.go.jp

We have carried out dual-comb spectroscopy and observed in a simultaneous acquisition a 140-THz-wide spectrum from 1.0 to 1.9 µm using two fiber-based frequency combs phase-locked to each other. This ultra-broad wavelength bandwidth is realized by setting the difference between the repetition rates of the two combs to 7.6 Hz using the sub-Hz-linewidth fiber combs. The recorded spectrum contains five vibration-rotation bands of $C_2H_2$, $CH_4$, and $H_2O$ at different wavelengths across the whole spectrum. The determined transition frequencies of $C_2H_2$ agree with those from the previous sub-Doppler resolution measurement of individual lines using CW lasers within 2 MHz.

Optical frequency combs have been expected to work as broadband light sources for advanced spectroscopy in addition to their use as optical frequency rulers for precise measurements. An optical frequency comb is a collection of lasers with identical frequency intervals, and molecular responses can be recorded in the amplitude and phase of each comb mode. To benefit from the advantage that these combs have over thermal light sources, individual comb modes must be separated in the spectrum. Several methods have been demonstrated for accomplishing this including the combination of an optical grating and a Fabry-Perot cavity,[1,2] a virtually imaged phased array (VIPA),[3] and Fourier transform spectroscopy (FTS).[4,5] Dual-comb spectroscopy is one such method, in which atomic or molecular absorption information is stored in each mode of the first comb (signal comb) and retrieved for the individual comb mode by the second comb (local comb) through multi-heterodyne detection.[6,7] Frequency accuracy and acquisition time superior to those of conventional FTS and a spectral resolution equal to the repetition rate of the signal comb have been demonstrated using dual-comb spectrometers based on mode-locked fiber lasers.[8–13]

In the dual-comb spectroscopy, optical pulses from signal and local combs with respective repetition rates of $f_{rep,S}$ and $f_{rep,L} = f_{rep,S} - \Delta f_{rep}$ interfere with each other, and the beat notes appear at a frequency of $N \cdot \Delta f_{rep}$ ($N$: integer) in the radio frequency (RF) region. The subscripts "S" and "L" indicate signal and local combs, respectively. In other words, the signal comb mode is mapped to the RF comb mode with a frequency reduction ratio of $\Delta f_{rep}/f_{rep,S}$. When the RF comb frequency of $N \cdot \Delta f_{rep}$ exceeds $f_{rep,L}/2$, more than two signal comb modes are mapped on the same RF comb mode (aliasing). To maintain a one-to-one correspondence between the signal and RF comb modes, the RF comb frequency is limited to a maximum value of $f_{rep,L}/2$ which corresponds to the optical frequency span of $f_{rep,S} f_{rep,L} /(2\Delta f_{rep})$ (Nyquist condition).[6,7] On the other hand, the $\Delta f_{rep}$ value must be sufficiently larger than the relative linewidth between the two combs; otherwise, the beat notes from the adjacent signal comb modes will overlap and thus cannot be resolved. Consequently, broadband dual-comb spectroscopy requires combs with a narrow relative linewidth.

Previously, a spectral bandwidth of 43 THz was observed in a simultaneous acquisition by setting $\Delta f_{rep}$ to 95 Hz using fiber combs with Hz-level relative linewidths.[8] Several approaches have been demonstrated for reducing the effective relative linewidth. For instance, relative phase fluctuations between the combs were corrected on the recorded interferogram in real time.[9] In another example, there was a fluctuation in delay time between two pulses from the free-running combs, but the sampling time interval for data acquisition was synchronized with the delay.[10] In this study, we demonstrate 140-THz-wide dual-comb spectroscopy that is achieved by employing two fiber-laser-based combs with a relative linewidth at the sub-Hz level.[14,15] The achieved spectral bandwidth in a simultaneous acquisition is the broadest one to the best of our knowledge. The observed spectrum contains rotation-vibration bands of acetylene ($C_2H_2$) at 1.04 and 1.52 μm, of methane ($CH_4$) at 1.67 μm, and of water ($H_2O$) in the atmosphere at 1.38 and 1.87 μm. We also determine the transition frequencies of $C_2H_2$ at 1.52 μm and compare them with those yielded by previous sub-Doppler resolution measurements[16] to validate the accuracy of the frequency measurement of the Doppler-broadened line center using the present dual-comb spectrometer.

Figure 1 shows the configuration of our dual-comb spectrometer. The signal and local combs employ an erbium-based mode-locked fiber laser as the comb oscillator. It has an electro optic modulator (EOM) for fast servo control[14,15] and a delay line for the tuning of $f_{rep}$ in the laser cavity. The repetition rates of the combs are about 48 MHz. The output of the individual combs is divided into three branches, amplified by erbium-doped fiber amplifiers, and then spectrally broadened with highly nonlinear fibers. The waves from the branches are used to detect the carrier-envelope offset beat ($f_{ceo}$), to detect the beat note between a CW reference laser and the nearest comb mode ($f_{beat}$), and to record the dual-comb spectrum. $f_{ceo}$ is phase-locked at a reference frequency. One of the individual comb modes is phase-locked to a 1.54-μm CW laser which is stabilized to an ultra-stable cavity. The error signal is fed back to the EOM, a piezoelectric transducer, and a Peltier element in the comb oscillator with different time constants. By virtue of the fast control of the cavity length with the EOMs, the relative linewidth between the combs falls to less than the sub-Hz level. This is confirmed by observing the beat note between the two combs with an RF spectrum analyzer. The narrow relative linewidth allows us in principle to set $\Delta f_{rep}$ to the Hz level. The observable spectral bandwidth is limited to 140 THz by the spectral response of InGaAs PIN photodiodes on the low frequency side and by the spectral coverage of the combs on the high frequency side. Here we adjust $\Delta f_{rep}$ to 7.6 Hz, which is appropriate to record a 140-THz wide spectrum.

The optical pulses from the signal comb are transmitted through a 50-cm-long absorption cell filled with 2.7-kPa $CH_4$ and a 15-cm-long 26-pass White cell of 2.6-kPa $C_2H_2$. The transmitted signal wave overlaps with the local wave at a polarization beam splitter. These waves are divided into two beams and detected with two InGaAs detectors for balanced detection. The detected signal is an interferogram of the local and signal waves, which is guided to a 14-bit digitizer through an anti-aliasing filter with a pass-band from 0.5 to 21.4 MHz, and is sampled at the repetition rate of the local comb.

To enhance the sensitivity by the accumulation of interferograms, we employ a coherent averaging technique.[17] Thus, we are able to accumulate interferograms coherently for a few seconds which is determined by the inverse of the sub-Hz-level relative linewidth of the combs. In addition, the following two real-time compensations are employed to realize a long-term accumulation. First, we correct the carrier phase drift of the interferograms caused by the fluctuations in optical pass length on a computer.[18] Second, $f_{beat,L}$ and $f_{ceo,L}$ are actively controlled to compensate the frequency drift of the CW laser (~ 9 kHz/day); otherwise, the drift induces a variation in repetition rate and breaks the conditions for coherent averaging.

Figure 2(a) shows an observed spectrum containing the $2\nu_3$ band of $CH_4$ at 1.67 μm, the $\nu_1 + \nu_3$ band of $C_2H_2$ at 1.52 μm, the $3\nu_3$ band of $C_2H_2$ at 1.04 μm, the $\nu_1 + \nu_3$ band of $H_2O$ at 1.38 μm, and the $2\nu_2 + \nu_3$ band of $H_2O$ at 1.87 μm. Figures 2(b)−(f) show the expanded views of these bands. Figure 2(g) is an expanded view of Fig. 2(f). The $\Delta f_{rep}$ value is 7.6 Hz and corresponds to a period of 130 ms, which is the minimum time needed for recording the whole spectrum with the data points of 6.4 million. The spectrum of Fig. 2(a) is Fourier-transformed from an interferogram averaged over 10,000 measurements of as long as 22 min. The horizontal axis is scaled with the absolute frequency calculated from the mode number determined from the frequency of the 1.54 μm CW laser, the carrier-envelope offset frequencies, and the repetition rates which are measured using a hydrogen maser. The frequency step of the spectrum is 48 MHz. This is the repetition rate of the signal comb and is regarded as the resolution of the spectrometer. It is worth noting here that, although the resolution of the spectrometer is limited by the repetition rate of the signal comb, the frequency uncertainty of the horizontal axis is limited by the frequency reference (the hydrogen maser) and is much less than 1 kHz. Therefore, it should be possible to determine the frequency of the observed lines with an uncertainty well below the resolution of the spectrometer. The rotation-vibration lines of the $\nu_1 + \nu_3$ band of $C_2H_2$ completely absorb the signal wave because the sample pressure is set so that it is adequate for the absorption lines at 1.04 μm, the absorption intensity of which is three orders of magnitude smaller than those at 1.52 μm.

Figure 3 shows a normalized absorption spectrum of $C_2H_2$ in the 1.52 μm band and a calculated spectrum derived from the line parameters of HITRAN2012[19] together with the discrepancies between them. The absorption spectrum is recorded with a sample pressure of 60 Pa, $\Delta f_{rep}$ of 33 Hz, and an averaging number of 10,000 over 5 min. A reference interferogram for normalization is recorded using an empty cell. The $P(25)$-to-$R(29)$ transitions and the intensity alternation are clearly observed in Fig. 3. In addition, relatively weak lines including the $\nu_1 + \nu_2 + \nu_4 + \nu_5$ band and some hot bands are also observed. The calculated spectrum is fitted to the observed spectrum with one adjustable parameter, namely absorption length. The residuals indicate that the observation agrees well with the calculation as regards both absorption intensity and transition frequency.

Figure 4 shows the absorption spectrum around the $R(11)$ line of the $\nu_1 + \nu_3$ band of $^{12}C_2H_2$. The spectrum is fitted to a Gauss function with the center frequency, amplitude, and linewidth as adjustable parameters. The determined center frequency of the $R(11)$ lines through the fitting is 197 343 962.5(0.7) MHz. Table I shows the determined transition frequencies ($\nu_{obs}$) of some of the typical lines in the band (P and R branch lines, each with even and odd $J$ values). Here, the $\sigma$ values are standard deviations of the fitting, and the $\delta = \nu_{obs} - \nu_{ref}$ are discrepancies with the previous sub-Doppler resolution data ($\nu_{ref}$).[16] The standard deviations are typically less than 2 MHz, which is limited by the signal-to-noise ratio (SNR) of the spectrum. The discrepancies are typically around 1 MHz, which originates in mainly residual fringes of the base line. The pressure shift of the $^{12}C_2H_2$ lines is estimated to be –0.13 MHz[20] and is negligible compared with the standard deviations.

Here, we discuss the SNR of broadband dual-comb spectroscopy. We estimate the noise level of our dual-comb spectrometer on the basis of the discussion in Ref. 21. The relative intensity noise (RIN) of the spectrally broadened comb is dominant in our measurement across 1.0−1.9 μm (Fig. 2). The measured RIN is about –116 dBc/Hz, which corresponds to a normalized noise spectral density of 17.8 Hz$^{-1/2}$. Therefore, the SNR is about 2.0 for 22 min average. In practice, the SNR of the $^{12}C_2H_2$ spectrum at 1.04 μm approximately ranges from 5 to 10. We expect that the SNR could be improved by stabilizing the comb power to reduce the RIN.

Our spectrometer is able to accumulate interferograms for a long period of time to obtain a required SNR. The inset of Fig. 2(g) shows a spectrum of the same lines recorded using 220,000 times averaging over 8 h. When we accumulate interferograms for 8 h, the SNR improved proportionally to the accumulation time, and significant spectral fringes of about 3 GHz appear, as shown in the inset of Fig. 2(g). We consider that these fringes originate from the spectrum of the comb oscillators and remain even after the spectral broadening in the highly nonlinear fiber. They become residual fringes on the base line of the normalized spectrum and cause a center–frequency shift. Therefore, comb oscillators with smooth spectra are required for the center-frequency determinations in such high-SNR spectroscopy.

In conclusion, we have developed an ultra-broadband dual-comb spectrometer, which simultaneously possesses a broad bandwidth, a high spectral resolution, and a high frequency accuracy. These excellent characteristics are superior to those of the conventional FTS and will stimulate various advanced applications such as the identification of the chirality of chiral molecules. In addition, a broader spectral coverage will offer new spectroscopic approaches to applications such as the multicomponent analysis and investigation of the relaxation associated with several vibrational states.[22] Combs with a sub-Hz level linewidth simply realize ultra-broadband dual-comb spectrometers, which will become ubiquitous tools in the near future.

**Acknowledgments**

This work is supported by MEXT/JSPS KAKENHI Grant Numbers 23244084, 23560048 and 22540415, Collaborative Research Based on Industrial Demand from the Japan Science and Technology Agency, and the Photon Frontier Network Program of the Ministry of Education, Culture, Sports, Science and Technology, Japan.

## Figure Captions

**Fig. 1.** Configuration of dual-comb spectrometer. EOM: electro-optical modulator; FFT: fast Fourier transform. A 50-cm-long single-pass cell is filled with 2.7 kPa $CH_4$, and a 15-cm-long 26-pass White cell is filled with 2.6 kPa $C_2H_2$.

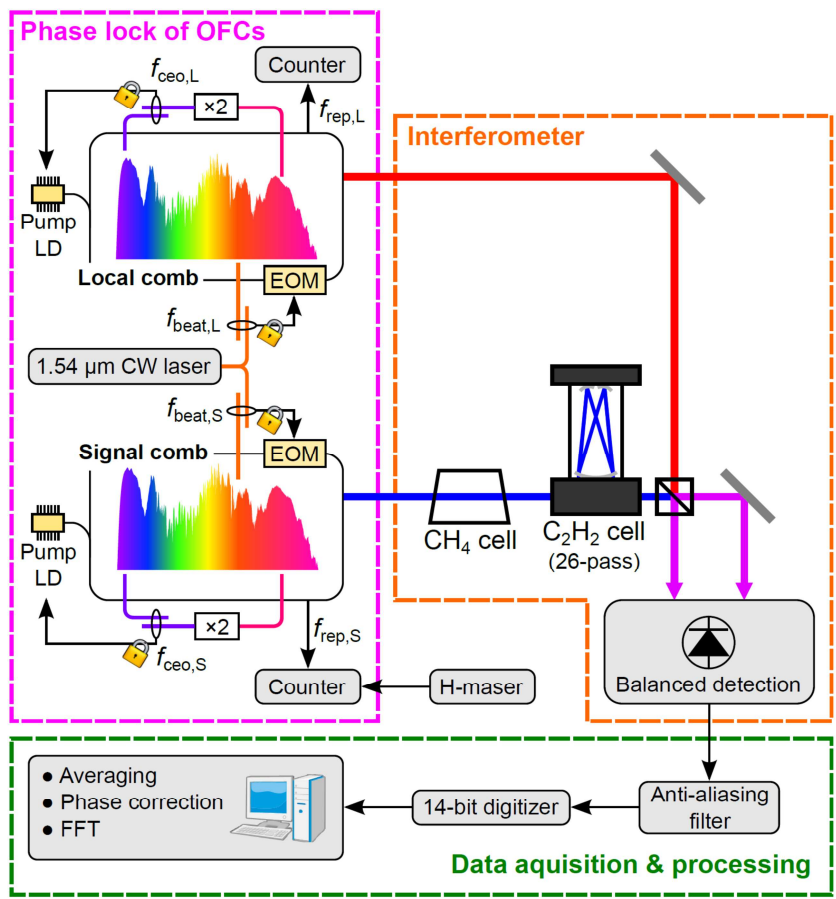

**Fig. 2.** Observed dual-comb spectrum. (a) Entire spectrum from 1.0 to 1.9 μm. The black structures represent the variation of the comb spectrum. (b)−(g) Expanded spectra of (a). (b) $H_2O$ $2\nu_2 + \nu_3$ band at 1.87 μm (160 THz). (c) $CH_4$ $2\nu_3$ band at 1.67 μm (180 THz). (d) $^{12}C_2H_2$ $\nu_1 + \nu_3$ band at 1.52 μm (197 THz). (e) $H_2O$ $\nu_1 + \nu_3$ band at 1.38 μm (218 THz). (f) $^{12}C_2H_2$ $3\nu_3$ band at 1.04 μm (289 THz). (g) Expanded spectrum of (f). The inset of (g) shows a spectrum of the same lines recorded using 220,000 times averaging over 8 h.

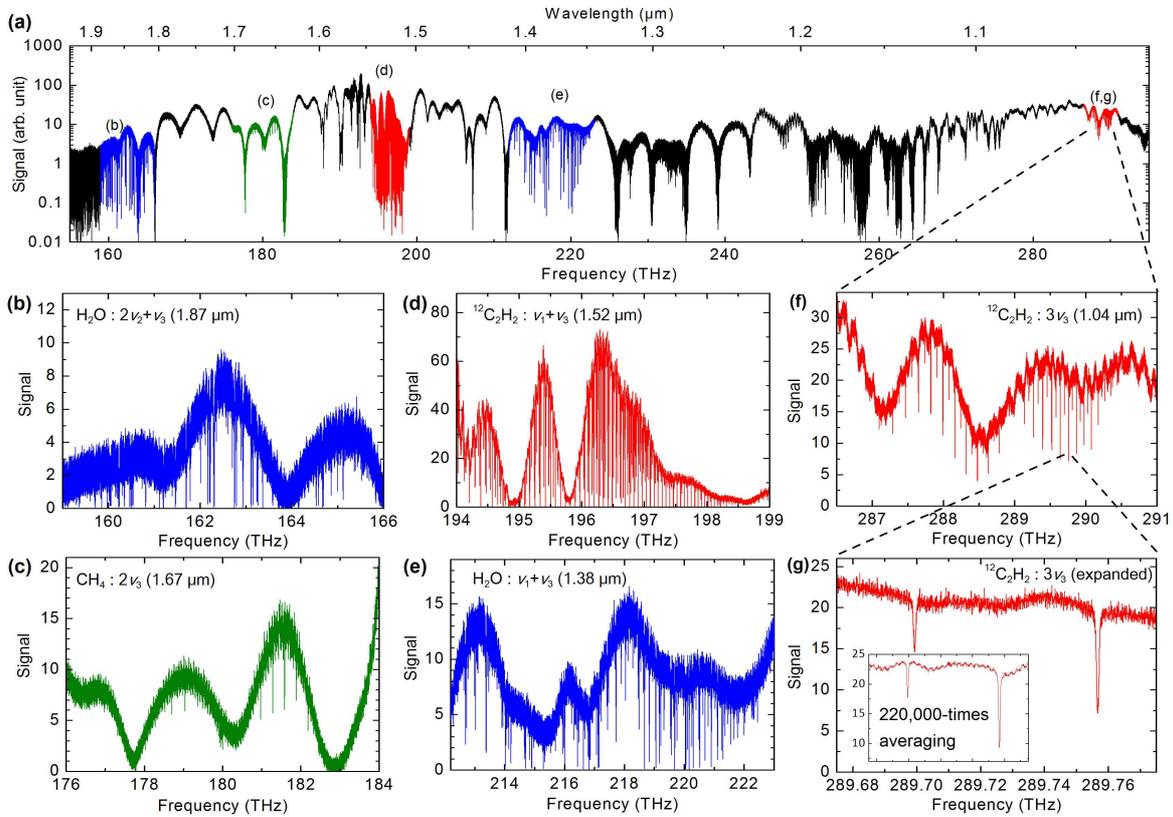

**Fig. 3.** Normalized (red) and calculated (blue) spectra and residuals between them (green) for the $\nu_1 + \nu_3$ band of $^{12}C_2H_2$ at 1.52 μm (197 THz). The sample pressure is 60 Pa.

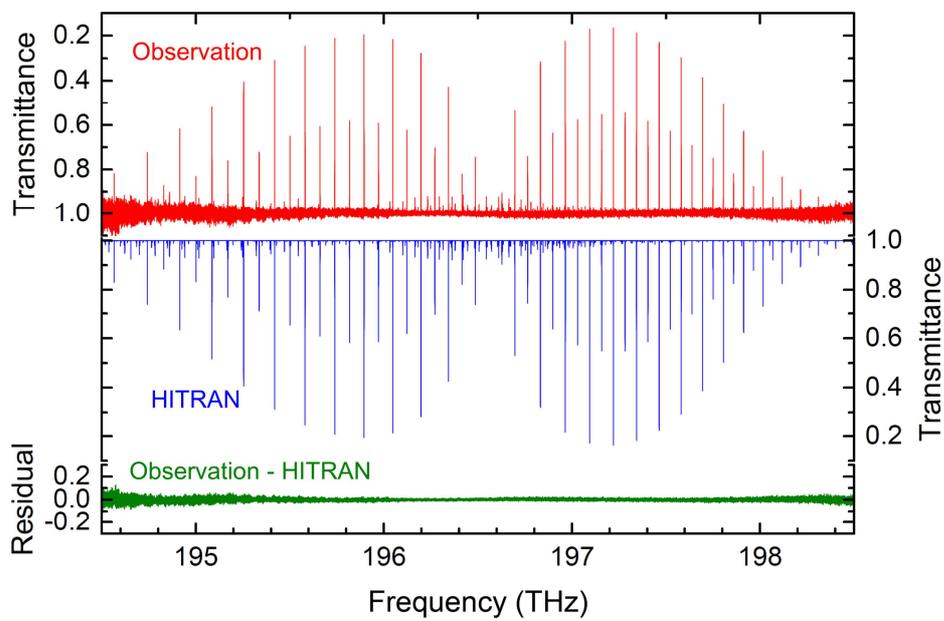

**Fig. 4.** Observed (black circles) and calculated (red curve) spectra and the residual between them (red circles) for the $R(11)$ line of the $\nu_1 + \nu_3$ band of $^{12}C_2H_2$.

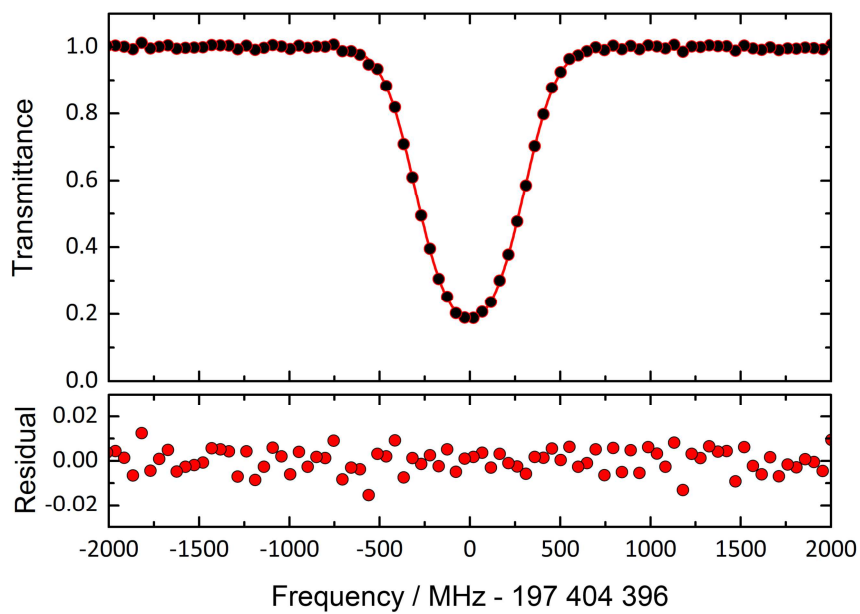

**Table I.** Frequency list of the $\nu_1 + \nu_3$ band of $^{12}C_2H_2$.

| Line | $\nu_{obs}$ (MHz) | $\sigma$ (MHz) | $\delta$ (MHz) |
|------|-------------------|----------------|----------------|
| P(12) | 195 660 692.5 | 2.0 | –0.3 |
| P(11) | 195 739 648.8 | 0.8 | –0.7 |
| R(11) | 197 343 962.5 | 0.7 | –0.1 |
| R(12) | 197 404 396.2 | 1.6 | 0.6 |